\documentclass{article}
\usepackage[utf8]{inputenc}
\usepackage{authblk}
\usepackage{dcolumn} 
\usepackage{bm}        
\usepackage{amssymb}
\usepackage[numbers]{natbib}
\usepackage{graphicx}
\usepackage{hyperref}

\title{\textbf{Quantum communication with photons}}

\author[1,2,3]{Mario Krenn}
\author[1,2]{Mehul Malik}
\author[1,2]{Thomas Scheidl}
\author[1,2]{Rupert Ursin}
\author[1,2,3]{Anton Zeilinger}
\affil[1]{\textit{\small{Institute for Quantum Optics and Quantum Information, Austrian Academy of Sciences, Boltzmanngasse 3, 1090 Vienna, Austria}}}
\affil[2]{\textit{\small{Vienna Center for Quantum Science and Technology, Faculty of Physics, University of Vienna, Boltzmanngasse 5, A-1090 Vienna, Austria}}}
\affil[3]{\small correspondence to: \href{mailto:mario.krenn@univie.ac.at}{mario.krenn@univie.ac.at} and \href{mailto:anton.zeilinger@univie.ac.at}{anton.zeilinger@univie.ac.at}}

\begin{document}

\maketitle

\begin{abstract}
The secure communication of information plays an ever increasing role in our society today. Classical methods of encryption inherently rely on the difficulty of solving a problem such as finding prime factors of large numbers and can, in principle, be cracked by a fast enough machine. The burgeoning field of quantum communication relies on the fundamental laws of physics to offer unconditional information security. Here we introduce the key concepts of quantum superposition and entanglement as well as the no-cloning theorem that form the basis of this field. Then, we review basic quantum communication schemes with single and entangled photons and discuss recent experimental progress in ground and space-based quantum communication. Finally, we discuss the emerging field of high-dimensional quantum communication, which promises increased data rates and higher levels of security than ever before. We discuss recent experiments that use the orbital angular momentum of photons for sharing large amounts of information in a secure fashion.
\end{abstract}

\newpage

\tableofcontents

\newpage

%%%%%%%%%%%% START READING HERE %%%%%%%%%%%%
\section{Introduction}

Ever since its inception, quantum physics has changed our understanding of the fundamental principles of nature. Apart from their impact on all fields of academic research, these insights have merged together with the field of information science to create the novel field of quantum information. Quantum information science provides qualitatively new concepts for communication, computation, and information processing, which are much more powerful than their classical counterparts. Quantum information is an intriguing example where purely fundamental and even philosophical research can lead to new technologies. The developments in this young field recently experienced a worldwide boom---as is evidenced by the increasing number of quantum information centers being founded in countries all over the world. Although its long-term industrial applications cannot be clearly anticipated, it is clear that quantum information science entails a huge potential economic impact. For reasons of space we limit ourselves to polarisation and orbital angular momentum (OAM) as information carrying degrees of freedom.

%\begin{figure}[h!]
%\centering
%\includegraphics[scale=0.5]{FOV_from_ISS.png}
%\caption{A vision: Global Quantum Communication via satellites.}
%\label{fig:univerise}
%\end{figure}

\subsection{The Quantum Bit}

In classical information and computation science, information is encoded in the most fundamental entity, the bit. Its two possible values \textbf{0} and \textbf{1} are physically realized in many ways, be it simply by mechanical means (as a switch), in solids by magnetic or ferroelectric domains (hard drives), or by light pulses (optical digital media). All of these methods have one thing in common---one state of the device mutually excludes the simultaneous presence of the other---the switch is either \textbf{on} or \textbf{off}.

\begin{figure}[h!]
\centering
\includegraphics[scale=0.5]{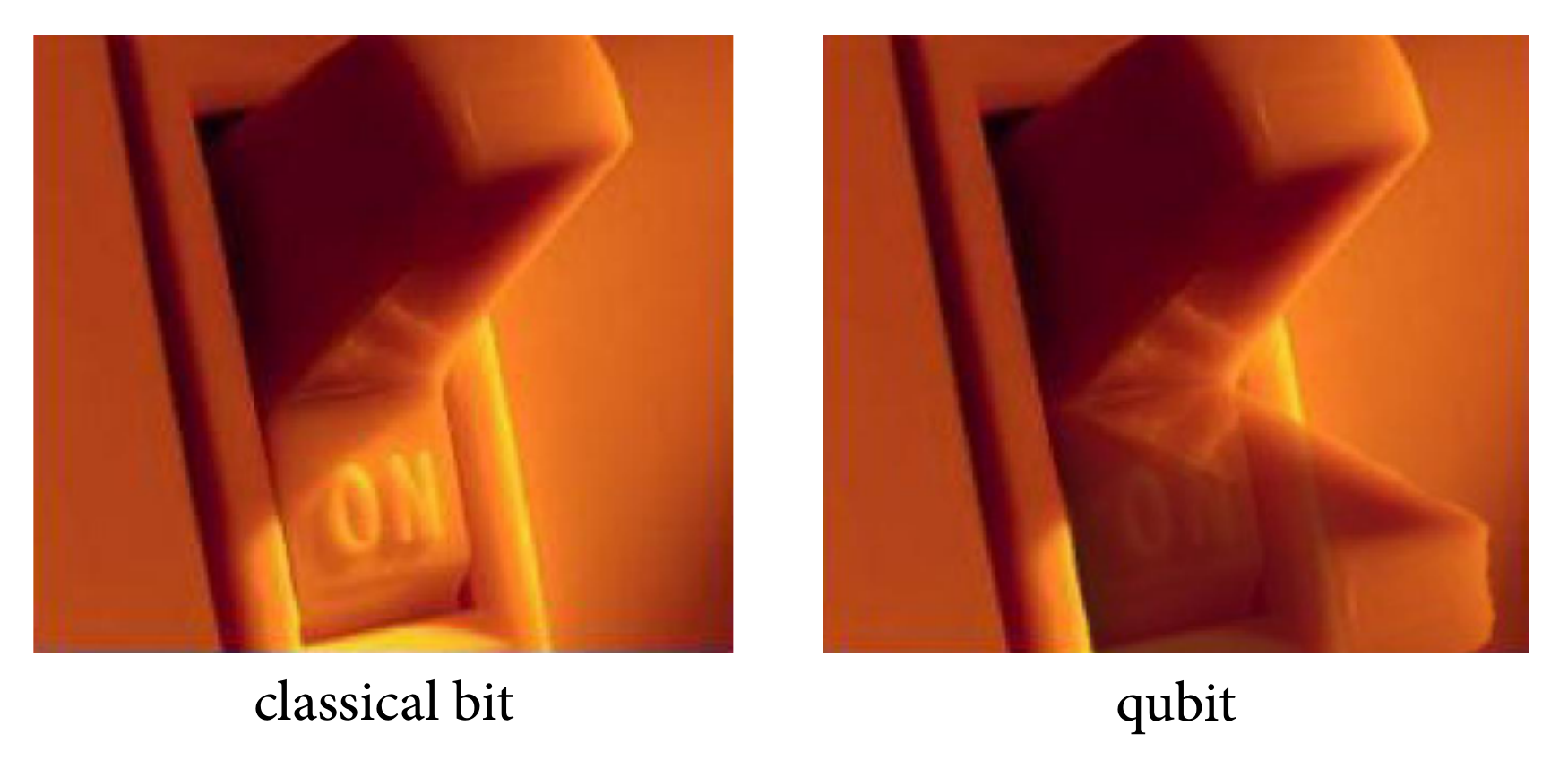}
\caption{An illustration of the difference between a classical bit and a qubit. The classical bit is always in a well-defined state while the qubit can also exist in a superposition of orthogonal states. (copyright University of Vienna)}
\label{fig:switch}
\end{figure}

The superposition principle entails one of the most fundamental aspects of quantum physics, namely to allow the description of a physical system as being in a probabilistic combination of its alternative states. This so-called \textit{superposition of states} not only provides all predictions for the outcome of a physical measurement, it also has drastic consequences for the nature of the physical state that we ascribe to a system. Its most important direct implication is the so-called \textit{no-cloning theorem}, which states that it is impossible to obtain a perfect copy of a qubit in an unknown state without destroying the information content of the original. The no-cloning theorem is the basis for the security of all quantum communication schemes described in the following sections, and will be explained later in more detail.

A qubit can be realized in many different physical systems such as atoms, ions, and super-conducting circuits. The most prominent physical realization of a qubit in view of a potential global-scale quantum communication network is with photons. Using photons, the two values of a bit, \textbf{0} and \textbf{1}, can be encoded in many different ways.
%For example, a photon can be in two different spatial modes such as the two output modes of a beam splitter, referred to as a spatial-mode qubit. Alternately, a photon can exist in two different temporal modes, for example when it has to traverse two paths of unequal lengths. This is commonly referred to as a time-bin qubit.
One possibility is to use two orthogonal polarisation states of a single photon, referred to as a polarisation qubit. In the latter case, one can ascribe the horizontal polarisation state of the photon with the logical value \textbf{0} and the vertical polarisation state with the value of \textbf{1}. Any arbitrary polarisation state can be obtained via a superposition of the horizontal and vertical state. The advantage of using photonic polarisation qubits is that they can be easily generated, controlled, and manipulated with rather simple linear optical devices like wave plates. Furthermore, since photons rarely exhibit interaction with the environment they are the best candidates for long-distance free-space transmission as would be required in a future network involving ground-to-space links.

To fully understand a qubit, it is important to distinguish between a coherent superposition and a mixture of possible states. For its use in quantum communication, it is important that a photon exists in a coherent superposition of its possible states. 
%For example, a photon that is in a coherent superposition of two spatial modes, such as the two output paths of a beam splitter, will exhibit interference when these two paths are recombined. A photon that is in a coherent superposition of two possible states has a fixed phase relationship between these states.%
For example, a polarisation qubit being in a coherent superposition of horizontal and vertical polarisations (with a certain phase relation) can be understood as a photon polarised diagonally at $+45^\circ$. A polarizer set at this angle will always transmit such a photon with 100\% probability (and zero probability when set to $-45^\circ$). However, a photon in a mixture (incoherent superposition) of horizontal and vertical polarisation states will be transmitted with 50\% probability.
\begin{figure}[h!]
\centering
\includegraphics[scale=0.5]{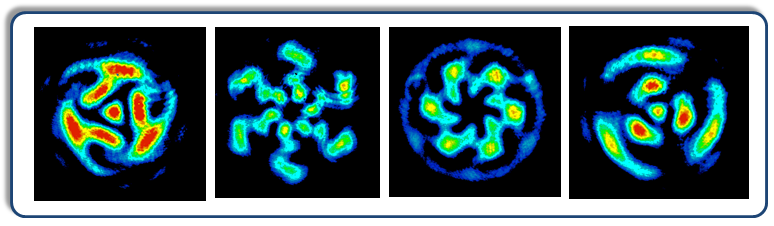}
\caption{Some types of higher-order spatial modes, which can carry more information than one bit per photon. (Image by Mario Krenn, copyright University of Vienna)}
\label{fig:HigherModes}
\end{figure}
Quantum superpositions, however, are not limited to just two possible states. The information carried by a photon is potentially enormous. While polarisation is necessarily a two-level (qubit) property, other degrees of freedom of a photon such as its spatial or temporal structure can have many orthogonal levels. For example, a photon can exist in a coherent superposition of different paths coming out of a multi-port beam splitter. These types of superpositions are referred to as ``high-dimensional" by virtue of their ability to encode large amounts of information. Consider a photon that is carrying a complicated image, such as that shown in Figure \ref{fig:HigherModes}. This image can be decomposed in terms of any orthonormal basis of spatial modes. The number of modes required for a complete description of this image dictate the number of levels, or dimensionality of this photon. One such basis is the set of Laguerre-Gaussian modes, which are described by a photon carrying a twisted wavefront. The phase structure of such a photon winds from 0 to $2\pi$ azimuthally around the optical axis, with the number of twists dictating the photon state dimensionality. Using such high-dimensional degrees of freedom of a photon for encoding surely increases the amount of information one can send per photon. However, a more subtle advantage of doing this is found in quantum communication---not only can one vastly increase the information capacity of quantum communication systems, one can also increase their security. This point is discussed in detail later in this chapter.\\

\subsection{Entanglement}
\begin{figure}[h!]
\centering
\includegraphics[scale=0.5]{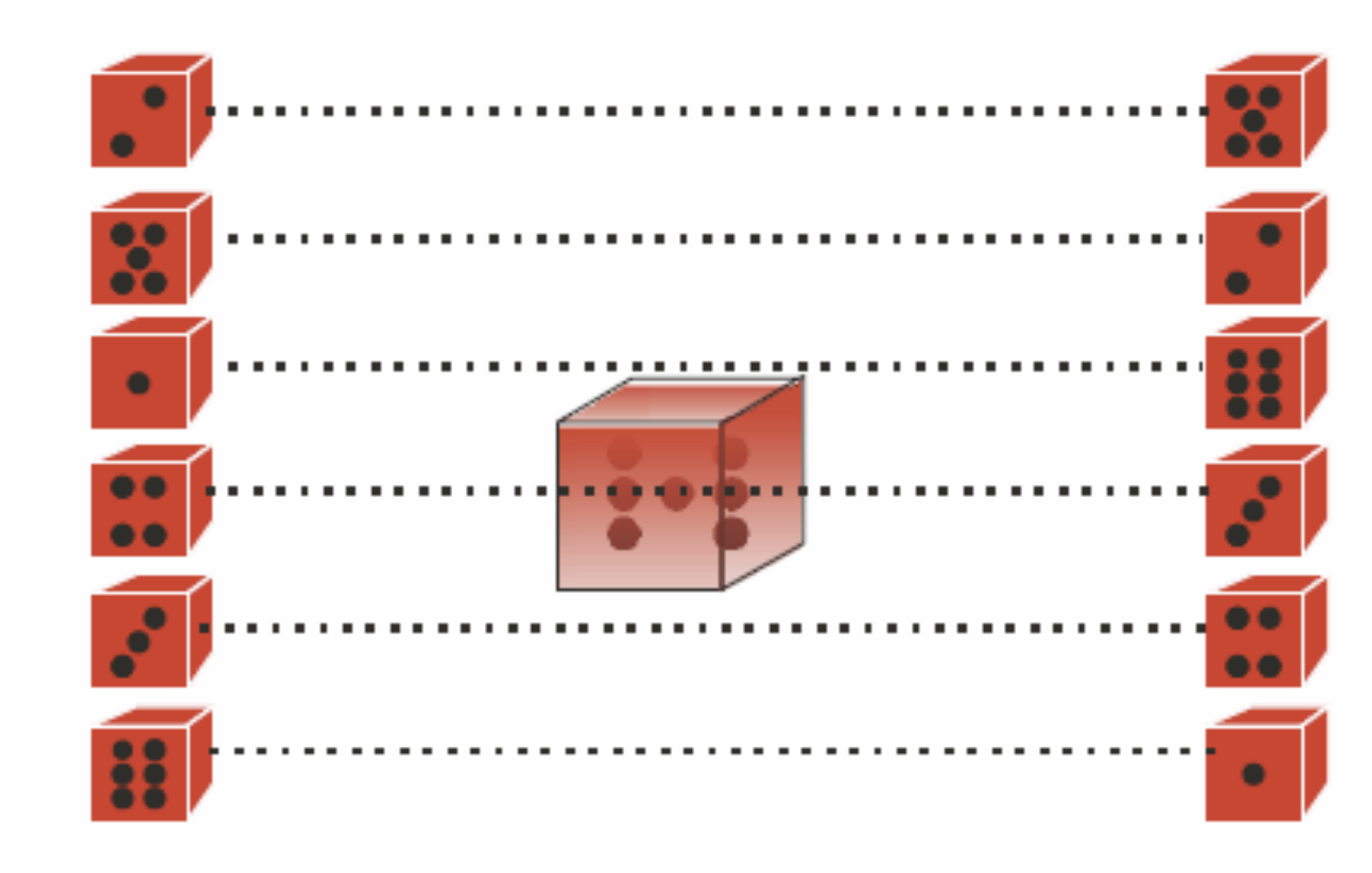}
\caption{If one could entangle a pair of dices with respect to their numbers, one can encode the message \textbf{7} by using their entanglement. None of the dices would carry this information on its own and a local measurement of the dice will result in a completely random result (without revealing the information). However, the results are perfectly correlated to add up to \textbf{7} for every joint measurement on the two dices. Note that a rolling dice corresponds to a six-dimensional qubit, which was prepared in a way unknown to us, and which is about to be measured in one out of six orthogonal bases.  (copyright University of Vienna)}
\label{fig:entanglement}
\end{figure}

The principle of superposition also holds for states containing several qubits. This allows for multi-qubit systems, which can only be described by joint properties. Such states are called \textit{entangled}, describing the fact that none of the particles involved can be described by an individual quantum state \cite{schrodinger1935gegenwartige,Einstein:1935hx,Bell:1964wu}. This is equivalent to the astonishing property of entangled quantum systems, that all of their information content is completely entailed in the correlations between the individual subsystems and none of the subsystems carry any information on their own. For example, when performing measurements on only one of two entangled qubits, the outcome will be perfectly random, i.e., it is impossible to obtain information about the entangled system. However, since the entangled state consists of two qubits, the correlations shared between them must consist of two bits of classical information. As a consequence, these two bits of information can only be obtained when the outcomes of the individual measurements on the separate subsystems are compared (see Figure \ref{fig:entanglement}).

Another intriguing feature of of entangled states is that a measurement on one of the entangled qubits instantaneously projects the other one onto the corresponding perfectly correlated state, thereby destroying the entanglement. Since these perfect correlations between entangled qubits are in theory independent of the distance between them, the entanglement is in conflict with the fundamental concepts of classical physics---locality (i.e.~distant events cannot interact faster than the speed of light) and realism (i.e.~each physical quantity that can be predicted with certainty corresponds to an ontological entity, a so-called ``element of reality") \cite{Bell:1964wu}. This has led to various philosophical debates about whether quantum mechanics can serve as a complete description of reality. However, there have been many experiments performed addressing this issue, and to date each of them has confirmed the predictions of quantum mechanics \cite{freedman1972experimental,aspect1982experimental,weihs1998violation,rowe2001experimental,scheidl2010violation}. One should note that while here we focus on polarisation and orbital-angular momentum entanglement, light can be entangled in its other degrees of freedom as well, such as time-frequency \cite{Franson:1989uu,Jha:2008wu} and position-momentum \cite{Klyshko:1988eea,Bennink:2004df,Leach:2012wb}.

\subsection{Mutually Unbiased Bases}
\begin{figure}[h!]
\centering
\includegraphics[scale=0.5]{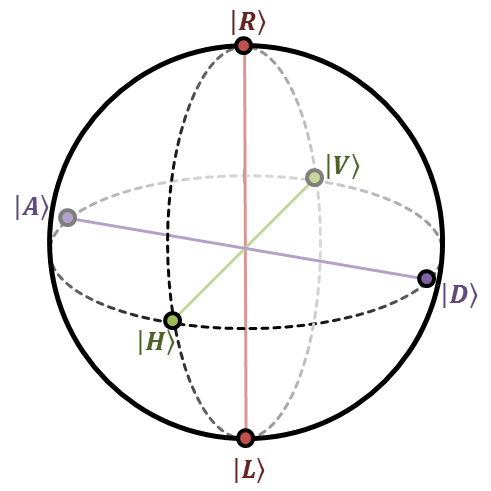}
\caption{The Bloch-sphere: Graphical representation of a two-dimensional qubit. There are three mutually unbiased bases---three ways of encoding information in different ways. In the case of polarisation, they correspond to horizontal and vertical (violet), diagonal and anti-diagonal (green) and right- and left-circular (red) polarisation. (Image by Mario Krenn, copyright University of Vienna)}
\label{fig:MUBs}
\end{figure}

One fascinating concept in quantum mechanics is the possibility to encode quantum information in different ways. In the simple example of the polarisation of light, there are three bases in which one can encode one bit of information (see Figure \ref{fig:MUBs}). These are the horizontal and vertical (H/V) basis, the diagonal and anti-diagonal (D/A) basis, and the left- and right-circular (L/R) basis. One can encode a bit in the H/V basis by considering 0 to be horizontal polarisation and 1 to be vertical polarisation. If a photon encoded in either H or V polarisation is measured in any of the other two bases, its information cannot be extracted. For example, in the case of measurements made in the D/A basis, in 50\% of the cases, a diagonally polarised photon will be observed; in the other cases the photon will be measured as anti-diagonally polarised. This property is the main ingredient for quantum cryptography, as we will see later. Furthermore, in higher-dimensional systems, fundamental properties of mutually unbiased bases are still open questions that are significant for quantum communication.

\subsection{Faster-than-light communication and the No-Cloning Theorem}
\begin{figure}[h!]
\centering
\includegraphics[scale=0.5]{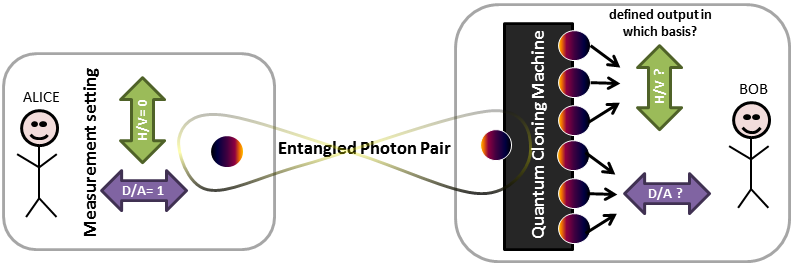}
\caption{Visualisation of a faster-than-light quantum communication protocol, if (!) quantum states could be cloned: Alice and Bob share an entangled photon pair. By choosing the measurement basis between horizontal/vertical or diagonal/antidiagonal polarisation, Alice projects the whole state into an eigenstate of that basis. This means that Bob's state is also defined in that basis. To find the basis chosen by Alice, Bob would need to measure more than one photon. If he could perfectly clone his photon, he could find the basis, and receive the information faster than light. Unfortunately, this is prohibited by the no-cloning theorem, a fundamental rule in quantum mechanics.  (Image by Mario Krenn, copyright University of Vienna)}
\label{fig:clone}
\end{figure}

As discussed above, two entangled photons are connected even though they can be spatially separated by hundreds of kilometers. The measurement of the first photon immediately defines the state of the second photon. Can one use that to transmit information faster than the speed of light?
If Alice and Bob share an entangled state and measure their respective photon in the same mutually unbiased basis (for instance, in the horizontal/vertical basis), they will always find the same result. However, whether they detect a horizontal or vertical photon is intrinsically random---there is no way that Alice could influence the outcome of Bob.
Regardless, there could exist a workaround, as shown in Figure \ref{fig:clone}. Alice could use her choice of measurement basis to convey information: either horizontal/vertical (H/V) if she wants to transmit 0 or diagonal/antidiagonal (D/A) if she wants to send 1. When she does this, Bob's photon is immediately defined in that specific basis. If Bob could now clone his photon, he could make several measurements in both bases and find out in which of the two bases his photon is well defined: If Alice measured in the H/V basis and finds an H outcome, all of Bob's measurements in the H/V basis will be H. However, his measurements in the D/A basis will show 50\% diagonal and 50\% antidiagonal. Thus, he knows that Alice has chosen the H/V basis, and thereby transmitted the bit value 0. 

Unfortunately, there is one problem with that protocol: It cannot exist. In 1982, Wootters and Zurek found that quantum mechanics forbids one to perfectly clone a quantum state \cite{Wootters:1982ex}. This profound result originates from a simple property of quantum mechanics, namely the linear superposition principle. We can inspect what a potential cloning-operation $\hat{C}$ would do. We use an input quantum state, and an undefined second photon $\left|X\right\rangle$. After the cloning operation, the second photon should have the polarisation property of the first photon. This is how our cloning machine would act on states in the H/V-basis:

\begin{eqnarray}
\hat{C}(\left|H\right\rangle\left|X\right\rangle)=\left|H\right\rangle\left|H\right\rangle \\
\hat{C}(\left|V\right\rangle\left|X\right\rangle)=\left|V\right\rangle\left|V\right\rangle
\end{eqnarray} 
The cloning-machine should work in every basis, thus we inspect what happens when we try to clone a diagonally polarised photon $\left|D\right\rangle$. Note that a diagonally polarised photon can be expressed in the H/V basis as a coherent superposition of a horizontal and a vertical part $\left|D\right\rangle=\frac{1}{\sqrt{2}}(\left|H\right\rangle + \left|V\right\rangle)$. The quantum cloning machine acts as 
\begin{eqnarray}
\hat{C}(\left|D\right\rangle\left|X\right\rangle)&=&\\\nonumber
&=&\hat{C}(\frac{1}{\sqrt{2}}(\left|H\right\rangle + \left|V\right\rangle)\left|X\right\rangle)\\\nonumber
&=&\frac{1}{\sqrt{2}}(\hat{C}\left|H\right\rangle\left|X\right\rangle + \hat{C}\left|V\right\rangle\left|X\right\rangle)\\\nonumber
&=&\frac{1}{\sqrt{2}}(\left|H\right\rangle\left|H\right\rangle + \left|V\right\rangle\left|V\right\rangle)
\end{eqnarray} 

The last line in equation (3) was obtained by using equations (1) and (2) for the cloning operator $\hat{C}$. The result is an entangled state that cannot be factorised into $\left|D\right\rangle\left|D\right\rangle$. If one were to measure either of the entangled photons individually, the result would be random, and certainly not $\left|D\right\rangle$. From this simple example it is clear that quantum cloning is not possible. This property prohibits faster-than-light communication, but it opens the door to many different quantum secret sharing protocols, such as quantum cryptography.

\subsection{Quantum Communication Schemes}
The counterintuitive quantum principles of superposition and entanglement are not only the basis of acquiring a deeper understanding of nature, but also enable new technologies that allow one to perform tasks which are not possible by classical means. When speaking about such ``quantum technologies", we refer to technologies that make explicit use of these kinds of quantum properties that do not have a classical analog.
Quantum information science and quantum communication are important ingredients in future quantum information processing technologies. They enable the transfer of a quantum state from one location to another. All quantum communication schemes have in common that two or more parties are connected via both a classical communication channel and a quantum channel (i.e. a channel over which quantum systems are transmitted). Typically, measurements are performed on the individual quantum (sub-) systems and the measurement bases used for every measurement are communicated via the classical channel. Here, we focus on quantum communication with discrete variables. However, we should mention that there exists a parallel branch of quantum communication that is based on continuous variables, where extensive theoretical and experimental work has been performed. More information on this field can be found in \cite{Weedbrook:2012fe} and references therein. 

\begin{figure}[h!]
\centering
\includegraphics[scale=0.5]{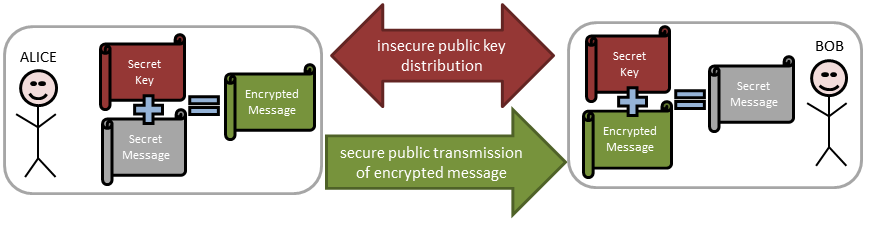}
\caption{Scheme of a classical symmetric cryptographical system. Alice wants to send a secret message to Bob. In order to do so, Alice and Bob have to share a secret key. With this key, they can distribute messages securely. The bottleneck is the distrbution of the key. This problem is solved by quantum cryptography.  (Image by Mario Krenn, copyright University of Vienna)}
\label{fig:symmCrypt}
\end{figure}

\subsection{Quantum Key Distribution}
If two parties want to share a secret message, they have two options: the first possibility is to share a random key that is the size of the message that needs to be encrypted with it (shown in Figure \ref{fig:symmCrypt}). The sender, let's call her Alice, performs a simple logical operator (an \textit{exclusive or}, XOR) of the message with the key, and gets the cipher. The cipher can only be read if the key is known. The receiver of the encrypted text, whom we will call Bob, can use the key to undo Alice's operation, which gives him the original message. The challenge lies in Alice and Bob having to share the entire secret key. 

\begin{figure}[h!]
\centering
\includegraphics[scale=0.5]{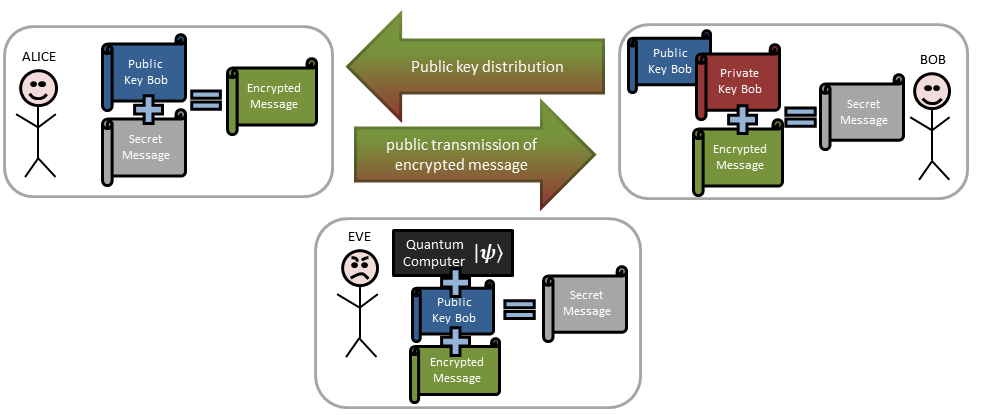}
\caption{Scheme of a classical asymmetric cryptographical system. Alice wants to send a secret message to Bob. In order to do so, Bob prepares a public and private key. Alice can then prepare an encrypted message for Bob with his public key. Usually, the message can only be decrypted by Bob with his private key. However, a powerful enough eavesdropper (for example, one with a quantum computer!) can infer Bob's private key from the public key, and can thus break the encryption protocol.  (Image by Mario Krenn, copyright University of Vienna)}
\label{fig:asymmCrypt}
\end{figure}

The alternative is a public-private key cryptography. This method, invented in the 1970s, is based on the computational complexity of finding the prime factors of large numbers. Again, Alice wants to send a secret message to Bob. Now Bob creates a pair of keys, a private and a public one. Everybody who has Bob's public key can encrypt messages for him. However, only Bob can decrypt those messages with his private key. However, it has been discovered by Peter Shor in 1994 that a quantum computer could factor prime numbers significantly faster than classical computers. It would allow an eavesdropper to read the secret message with only the information that is distributed publicly (see Figure \ref{fig:asymmCrypt}). One possible way to circumvent this problem is quantum key distribution. 

\begin{figure}[h!]
\centering
\includegraphics[scale=0.5]{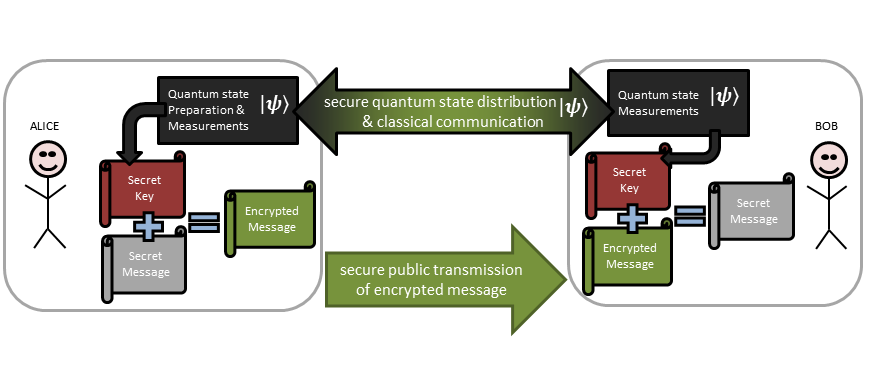}
\caption{Scheme of a quantum cryptographical system. Alice wants to send a secret message to Bob. In order to do so, a secret key is established over public (quantum) channels. Alice prepares a quantum state and transmits it to Bob. By making appropriate measurements, Alice and Bob can obtain a shared secret key. Alice then encrypts the message with this key and sends it to Bob; Bob can decrypt it with his copy of the key. Eavesdropping attempts during the key transmission appear as errors in the measurement results, allowing the presence of an eavesdropper to be detected. (Image by Mario Krenn, copyright University of Vienna)}
\label{fig:qkdcrypt}
\end{figure}

Quantum Key Distribution (QKD) allows two authorized parties to establish a secret key at a distance. The generation of this secret key is based on the same quantum physical principles that a quantum computer relies on. In contrast to classical cryptography, QKD does not simply rely on the difficulty of solving a mathematical problem (such as finding the prime powers of a large number). Therefore, even a quantum computer could not break the key. QKD consists of two phases (see Figure \ref{fig:qkdcrypt}. In the first phase the two communicating parties, usually called Alice and Bob, exchange quantum signals over the quantum channel and perform measurements, obtaining a raw key (i.e., two strongly correlated but non-identical and only partly secret strings). In the second phase, Alice and Bob use the classical channel to perform an interactive post-processing protocol, which allows them to distill two identical and completely secret (known only to themselves) strings, which are two identical copies of the generated secret key.
The classical channel in this protocol needs to be authenticated: this means that Alice and Bob identify themselves; a third person can listen to the conversation but cannot participate in it. The quantum channel, however, is open to any possible manipulation from a third person. Specifically, the task of Alice and Bob is to guarantee security against an adversarial eavesdropper, usually called Eve, tapping on the quantum channel and listening to the exchanges on the classical channel.

In this context security explicitly means that a non-secret key is never used: either the authorized parties can indeed create a secret key, or they abort the protocol. Therefore, after the transmission of the quantum signals, Alice and Bob must estimate how much information about raw keys has leaked out to Eve. Such an estimate is obviously impossible in classical communication: if someone is tapping on a telephone line, or when Eve listens to the exchanges on the classical channel, the communication goes on unmodified. This is where quantum physics plays a crucial role: in a quantum channel, leakage of information is quantitatively related to a degradation of the communication.
The origin of security of QKD can be traced back to the fundamental quantum physical principles of superposition and no-cloning. If Eve wants to extract some information from the quantum states, this is a generalized form of measurement, which will usually modify the state of the system. Alternatively, if Eve's goal is to have a perfect copy of the state that Alice sends to Bob, she will fail due to the no-cloning theorem, which states that one cannot duplicate an unknown quantum state while keeping the original intact.
In summary, the fact that security can be based on general principles of physics allows for unconditional security, i.e. the possibility of guaranteeing security without imposing any restriction on the power of the eavesdropper.

\begin{figure}[h!]
\centering
\includegraphics[scale=0.3]{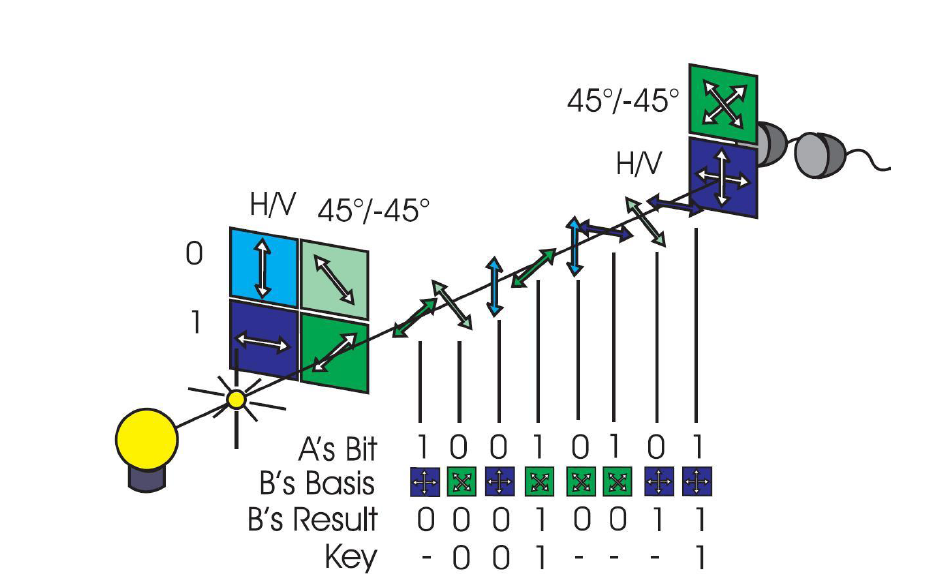}
\caption{An illustration of the coherent state BB84 protocol. Alice sends polarised single photons, prepared randomly in either of two complementary bases. Bob measures them, again randomly in one of the two bases. After publicly announcing their choice of bases, they obtain the sifted key from their data. (Copyright University of Vienna)}
\label{fig:bb84_protocol}
\end{figure}

The first Quantum Cryptography scheme was published by Bennett and Brassard in 1984 \cite{Bennett:1984wv} and is known today as the BB84 protocol. It requires four different qubit states that form two complementary bases (i.e. if the result of a measurement can be predicted with certainty in one of the two bases, it is completely undetermined in the other). These states are usually realized with four linear polarisation states of a photon forming two complimentary bases, for e.g. horizontal (H), vertical (V), diagonal (D) and anti-diagonal (A). As illustrated in Figure \ref{fig:bb84_protocol}, Alice sends single photons to Bob, which were prepared randomly in any of the four polarisation states and records the state of any sent photon. Bob receives and analyzes them with a two-channel analyzer, again randomly in one of the two complementary bases H/V or D/A. He records his measurement results together with the corresponding measurement basis. After enough photons have been transmitted, Bob communicates publicly with Alice and tells her which photons actually arrived and in which basis it was measured, but does not reveal the measurement result. In return, Alice tells Bob when she has used the same bases to prepare them, because only in these cases Bob obtains the correct result. Assigning the binary value \textbf{0} to H and D and the value \textbf{1} to V and A, leaves Alice and Bob with an identical set of \textbf{0}s and \textbf{1}s. This set is called the sifted key.

The security of the key distribution is based on the fact that a measurement of an unknown quantum system will (in most cases) disturb the system: If Alice's and Bob's sifted keys are perfectly correlated (which can be proven by comparing a small subset of the whole sifted key via classical communication), no eavesdropper tried to listen to the transmission and the key can be used for encoding a confidential message using the one-time pad (i.e., a specific key is exactly as long as the message to be encrypted and this key is only used once). In practical systems, however, there will always be some inherent noise due to dark counts in the detectors and transmission errors. As it cannot be distinguished whether the errors in the sifted key come from noise in the quantum channel or from eavesdropping activity, they all must be attributed to an eavesdropping attack. If the error is below a certain threshold Alice and Bob can still distill a final secret key using classical protocols for error correction and privacy amplification. If the error is above the threshold, the key is discarded and a new distribution has to be started. 

In contrast to the \textit{single-photon} protocols described above, entanglement based QKD uses entangled photon pairs to establish the secure key \cite{Ekert:1991kl,Bennett:1992cg}.  
Lets assume that Alice and Bob share a polarisation entangled two-photon state. Due to the perfect polarisation correlations between entangled photons, Alice and Bob will always obtain the same result, when they measure the polarisation state of their photon in the same measurement basis. Since both measure randomly in one of two complementary bases (just as in the BB84 protocol), they have to publicly communicate after they have finished their measurements, which photons they actually detected and in which basis it was measured. Again, they discard those results in which they disagreed in the measurement basis and finally end up with an identical set of \textbf{0}s and \textbf{1}s - the sifted key. Just as in the BB84 protocol, Alice and Bob authenticate their keys by openly comparing (via classical communication) a small subset of their keys and evaluating the bit error rate.

There are two big advantages in using entangled photons for implementing the QKD protocol. First, the randomness of the individual measurement results is intrinsic to the entangled state and therefore the randomness of the final key is ascertained. Second, an eavesdropper cannot mimic an entangled state by sending single photons in correlated polarisation states simultaneously to Alice and Bob. Hence, when using a subset of the transmitted photon pairs to examine the entanglement between them, secure communication is possible even though the operator of the entangled photon source might not be trustworthy.

\subsection{Quantum Teleportation}
\begin{figure}[h!]
\centering
\includegraphics[scale=0.7]{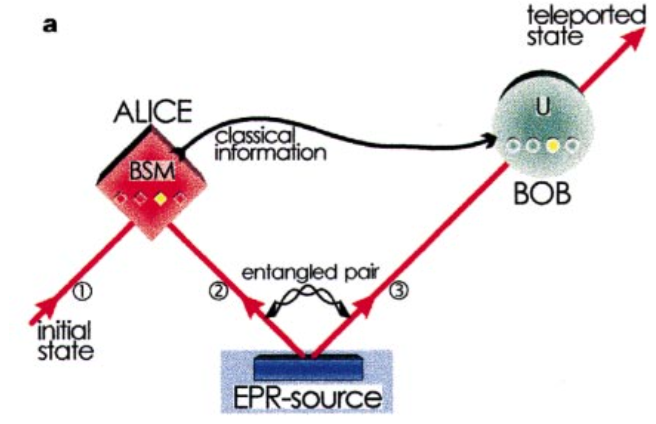}
\caption{Quantum state teleportation scheme. Picture taken from \cite{bouwmeester1997experimental}.}
\label{fig:teleport_scheme}
\end{figure}

Quantum teleportation is a process by which the state of a quantum system is transferred onto another distant quantum system without ever existing at any location in between \cite{Bennett:1993jc}. In contrast to what is often wrongly stated, this does not even in principle allow for faster-than-light communication or transport of matter. This becomes clearer when considering the entire three-step protocol of quantum teleportation (an illustration is shown in Figure \ref{fig:teleport_scheme}).

First, it is necessary that Alice (the sender) and Bob (the receiver) share a pair of entangled qubits (qubits 2 and 3 in the figure). Next, Alice is provided with a third qubit (qubit 1), the state of which she wants to teleport and which is unknown to her. In the last step, Alice destroys any information about the state of qubit 1 by performing a so-called Bell-state measurement (BSM) between qubits 1 and 2. As a consequence of this measurement and due to the initial entanglement between qubit 2 and 3, qubit 3 is instantaneously projected onto the same state as qubit 1. However, the teleportation protocol only works in cases, where the BSM resulted in exactly one out of four possible random outcomes. As a consequence, Bob needs to be notified by Alice about the outcome of the BSM in order to being able to identify the successful teleportation events. This requires classical communication between Alice and Bob and essentially limits the speed of information transfer within the teleportation protocol to the speed of the classical communication channel.

Quantum teleportation is an essential prerequisite for a so-called quantum repeater. A quantum repeater will be an important building block in a future network, since it allows to interconnect different network nodes. In a quantum repeater, two particles of independent entangled pairs are combined within a BSM, such that the entanglement is relayed onto the remaining two particles. This process is called entanglement swapping and will eventually allow to overcome any distance limitations in a global-scale network. However, in order to efficiently execute entanglement swapping, it has to be supplemented with an entanglement purification step requiring quantum memories.

\section{Long distance quantum communication}
\subsection{Ground-based long-distance experiments}
Quantum physics was invented to describe nature at the microscopic level of atoms and light. It remains an open question to what extent these laws are applicable in the macroscopic domain. In this respect, numerous ongoing research efforts pursue the goal of extending the distance between entangled quantum systems. They aim at investigating whether there are any possible fundamental limitations to quantum entanglement and if it is feasible to establish a global-scale quantum communication network in the future. In the past years, several free-space quantum communication experiments have been performed by several groups over various distances \cite{aspelmeyer2003long,resch2005distributing, ursin2007entanglement,schmitt2007experimental,peng2005experimental,yin2012quantum}, studying the feasibility of different quantum communication protocols over large distances. Starting with fairly short free-space links in the order of a few kilometers, the range was quickly extended up to today's world-record distance of 144 km, held by the authors of this article.

\begin{figure}[h!]
\centering
\includegraphics[scale=0.3]{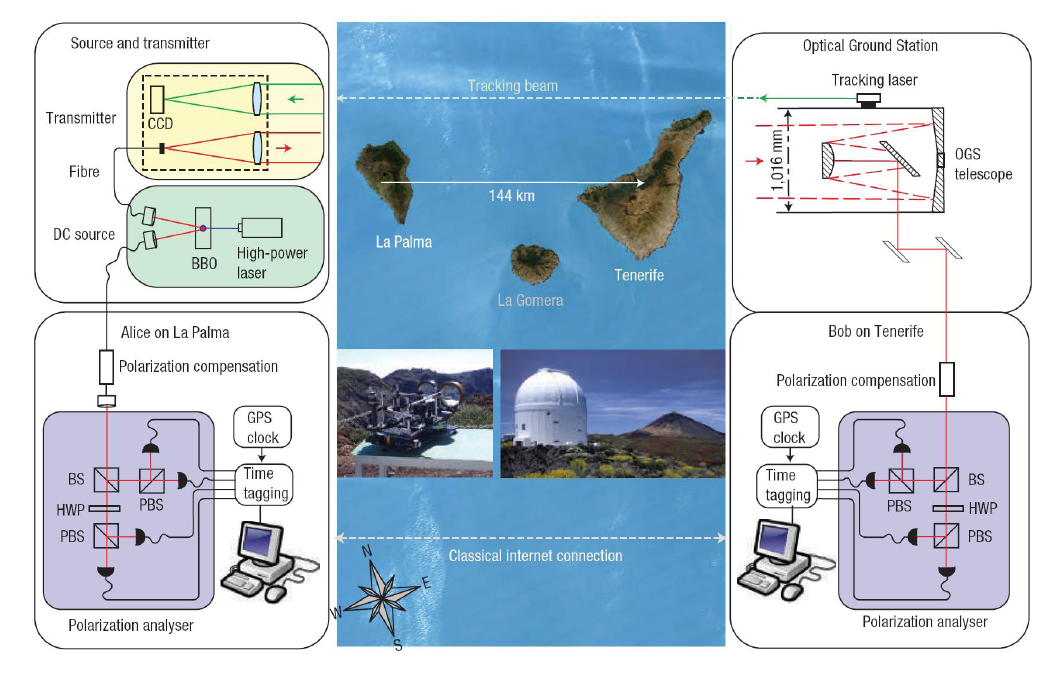}
\caption{An illustration of the experimental setup in the inter-island experiment from Ursin \textit{et al.}, distributing entangled photons over 144km between La Palma and Tenerife. Figure taken from \cite{ursin2007entanglement}.}
\label{fig:Fig_Canary05}
\end{figure}

One of the first experiments using a 144 km free-space link between the Canary Islands of La Palma and Tenerife was performed by Ursin \textit{et al.} in 2007 \cite{ursin2007entanglement}. In this experiment (see Figure \ref{fig:Fig_Canary05}), a source of entangled photon pairs was installed in La Palma at the top of the vulcano mountain Roque de los Muchachos at an altitude of 2400m.
One of the photons of an entangled pair was detected locally, while the other photon was sent to Tenerife. There, the optical ground station (OGS) of the European Space Agency (ESA), located at the Observatory del Teide at an altitude of 2400m, was used as the receiving telescope for the photons coming from La Palma. After analyzing the polarisation correlations between the associated photons on both islands, the scientists could verify that the photons are still entangled even though they have been separated by 144km. Additionally, the same group implemented quantum key distribution protocols based on both entangled as well as single photons \cite{ursin2007entanglement,schmitt2007experimental}. On the one hand, the results of these experiments addressed a question of fundamental physical interest, that entanglement can survive global-scale separations between the entangled particles. On the other hand, it verified that the OGS in Tenerife, which was originally built for laser communication with satellites, is also suitable to faithfully receive entangled photons. In combination, these results demonstrate the general feasibility for potential future space-based quantum communication experiments, thus setting the cornerstone for fundamental physical research as well as for potential applications of quantum mechanical principles in future network scenarios.

The achievements of these experiments were based on a combination of advanced techniques, laying the cornerstone for the Austrian researchers for a whole range of continuative activities employing the same free-space link between La Palma and Tenerife. In 2008, Fedrizzi \textit{et al.} \cite{fedrizzi2009high} generated entangled photon pairs in La Palma and sent both photons to Tenerife. The authors could verify entanglement between the photons detected in Tenerife and also implemented an entanglement based QKD protocol. This experiment was an important step towards a potential future quantum communication network, because with respect to the transmission loss, their experimental configuration was equivalent to a basic future network scenario, where entangled pairs are transmitted from a satellite to two separate receiving stations on ground

The long-distance experiments of our group so far involved only two photons. However, quantum communication protocols like teleportation or entanglement swapping, as described earlier, require more than two photons and will be of utmost important in a future network. Its experimental implementation, however, is substantially more complex than the two-photon protocols, necessitating a step back regarding the communication distance.
\begin{figure}[h!]
\centering
\includegraphics[scale=0.4]{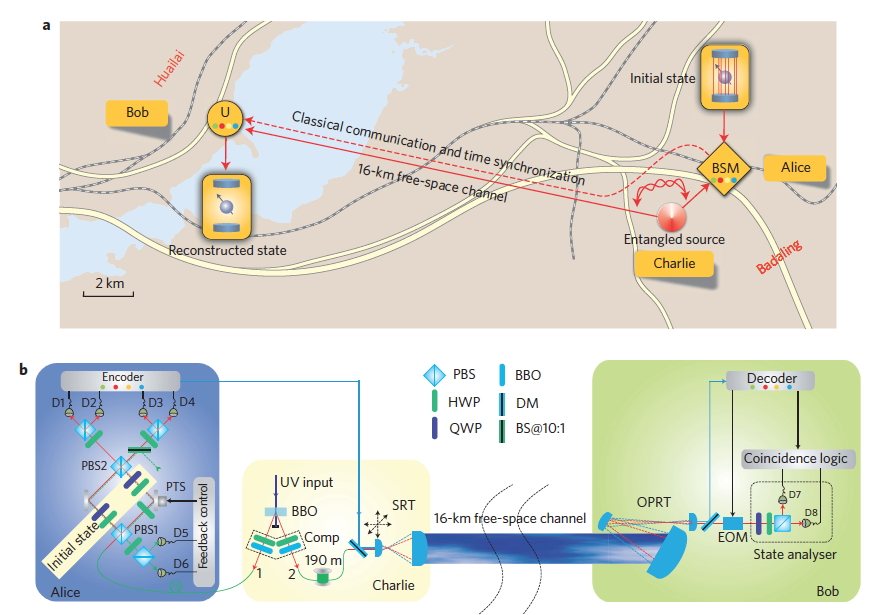}
\caption{A birds-eye view of the 16-km free-space quantum teleportation experiment of the Chinese group. Figure taken from \cite{jin2010experimental}.}
\label{fig:Fig_Teleport_Qinghai}
\end{figure}
In 2010, a group of Chinese researcher were the first to report on a long-distance free-space quantum teleportation experiment \cite{jin2010experimental}, demonstrating this protocol outside the shielded laboratory environment. They implemented a variant of the teleportation scheme described earlier and teleported the quantum states of photons over a distance of 16km. This achievement triggered a race between the Austrian and Chinese groups to push the distance record for teleportation even further. It lasted until 2012 that the Chinese group reported on a successful demonstration of quantum teleportation over a 97km free-space link across the Qinghai lake \cite{yin2012quantum}. But it was only 8 days later that also the Austrian group with the results of their work on long-distance quantum teleportation between La Palma and Tenerife, reporting a new distance record of 143 km \cite{ma2012quantum}.

The communication distances spanned in these experiments was in fact more challenging than expected for a satellite-to-ground link and thus the results of both groups proof the feasibility of quantum repeaters in a future space- and ground-based worldwide quantum internet. Together with a reliable quantum memory, these results set the benchmark for an efficient quantum repeater at the heart of a global quantum-communication network.

\begin{figure}[h!]
\centering
\includegraphics[scale=0.35]{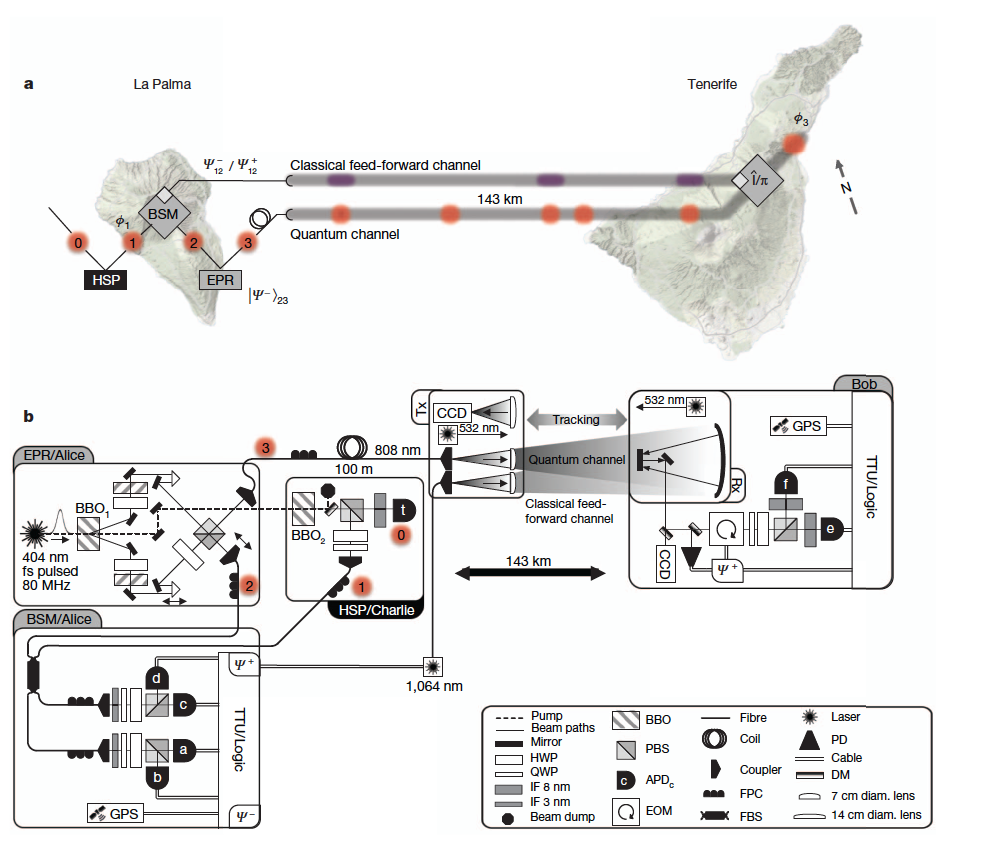}
\caption{An illustration of the experimental teleportation setup of the Austrian group, conducted at the Canary Islands. Figure taken from \cite{ma2012quantum}.}
\label{fig:Fig_Teleport_Canary}
\end{figure}

\subsection{Space-based quantum communication}
The experiments described above represent the state-of-the-art of long-distance quantum communication. Significantly longer distances are no longer possible on ground, since the curvature of the earth will then prevent direct line of sight links. The logical next step is to bring quantum technology into space and several international research initiatives in Europe, Singapore, China, USA, and Canada are currently pursuing related projects. 

It is a clear vision of the science community to establish a worldwide quantum communication network with all the advantages over it's classical counterpart described above. That requires significantly expanding the distances for distributing quantum systems beyond the capabilities of terrestrial experiments and can only be realized by tackling the additional challenge of bringing the concepts and technologies of quantum physics to a space environment. Long-distance quantum communication experiments have been underway for some time sending single photons through long optical fibers. The first scientific demonstration, still in the shielded laboratory, where conducted in the late 1990ties. The question to be answered at that time was, if the peculiar and fragile laboratory experiments can also be executed facing harsh real-world environmental conditions as are present in optical telecommunication networks.

There are limitations for high-speed quantum communication in optical fibers. For example, the maximum speed of generating, preparing and detecting single photons is on the order of a few Mbit per second using state-of-the-art high speed electronics. Due to the combination of noise in real detector-devices and transmission loss in the optical fiber, the distance, over which quantum information can be communicated is restricted to a few 100 km \cite{waks2002security}. Hence, for bridging distances on a global scale using optical fiber networks, the implementation of so-called quantum repeaters is paramout. Quantum repeaters are the quantum analog to classical optical amplifiers making global fibre communication as of today yet feasible. Quantum repeaters are a theoretical concept proposed in 1998 \cite{briegel1998quantum} and require as basic building blocks the concepts of quantum teleportation and quantum memories. Specifically, the combination of both is highly complex from a technological point of view, such that the development of a quantum repeater is yet in the early stages. The second solution to bridge distances on a global scale is to use satellite-to-earth and inter-satellite optical free-space connections \cite{ursin2007entanglement}. 

\begin{figure}[h!]
\centering
\includegraphics[scale=0.5]{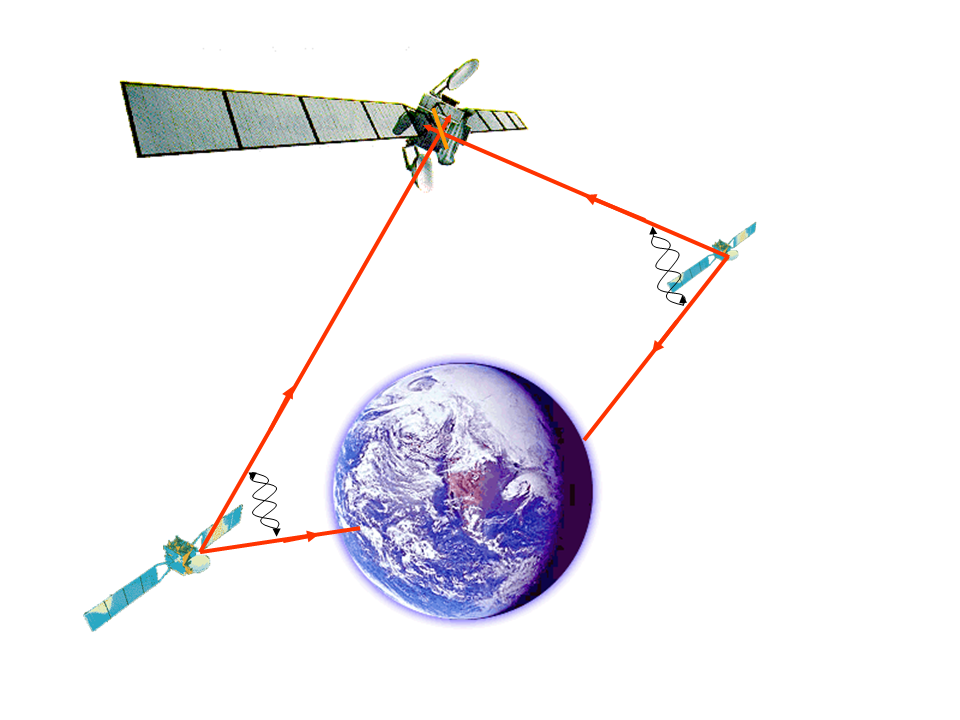}
\caption{A vision: Global Quantum Communication via satellites connecting any point on ground requiring optical ground station (taken from \cite{ursin2009space}).}
\label{fig:network}
\end{figure}

Figure \ref{fig:network} depicts a typical space-mission scenario for the distribution of entanglement from a transmitter terminal to two receiver stations (Alice and Bob). The quantum source installed on the transmitter emits pairs of photons in a desired entangled state. The photon pairs exhibit strong correlations in time, and entanglement in the degree of freedom in which the quantum information is encoded. The single photons comprising each of these entangled pairs are sent to Alice and Bob via free-space communications links (quantum links) established between the satellites or satellites and an optical ground station. The photons are collected via telescopes at the receiver terminals, where Alice and Bob each perform quantum measurements on their respective photons. Before initializing the transfer of information, the transmitter must establish a separate standard communications channel with Alice and Bob. This classical communications channel is subsequently used to send information about which basis state the measurements were performed on a given pair. The detection time of every arriving photon is recorded using fast single-photon detectors, and detection events that comprise an entangled pair are identified by means of their temporal correlations. The identification of photon pairs by their detection times requires the transmitter and receiver modules to establish and maintain a synchronized time basis, which can be achieved using an external reference, or autonomously via the classical communications link. Once the pair-detection events have been identified, Alice and Bob can reveal their stronger-than-classical correlations by communicating the bases of the quantum measurements performed on each photon pair via the classical communications channel. 

Distributing entangled photon pairs over long-distance links and revealing their quantum correlations is an immensely challenging task from a technological point of view, in particular due to the fact that, as a result of unavoidable losses in the quantum link, only a fraction of the photons emitted by the transmitter actually arrive at the receiver modules. The main sources contributing to losses along the optical transmission channel are atmospheric absorption and scattering, on the one hand, and diffraction, telescope pointing errors, and atmospheric turbulence, which all lead to beam broadening and thus limit the fraction of photons collected by the receiver aperture, on the other. Typical losses in such scenarios are in the order of -30 to -40 dB.

Nevertheless, in order to achieve feasible pair-detection rates at such huge link losses requires a very bright source of entangled photon pairs as well as minimizing losses in the transmission channel and the receivers. Note that, since correlated photon pairs are identified by their arrival times, there is an upper limit to how effective the photon production rate can mitigate against link loss. Once the time between two successive pair emissions at the source decreases below the timing jitter of the detectors, these two successive photons can no longer be distinguished from each other, such that as a result the quantum bit error ratio (QBER) will be increased. 

The pairs detected by the two terminals will ultimately comprise of photons steaming from the entangled photons source (the signal) but also from unavoidable sources of uncorrelated background photons (the noise). The background is from stray light the detector might see and the intrinsic dark counts of the photon avalanche detectors in use. The background can be mitigated to a certain extent bu using very narrow-band filters, allowing only those photons to be guided to the detector, who are at the wavelength of the quantum source in use. Also the common timing of the entangled photons are useful to mitigate noise pair counts.

Entangled photon sources maintaining both their high brightness and the quality of the emitted quantum state will have to be manufactured in a very reliable and stable manner to survive the launch of the satellite as well as the harsh space environment (radiation). First research and development projects funded by the European Space Agency were dedicated to the non-linear periodically poled Potassium Titanyl Phosphate (ppKTP) crystal, which is used in state-of-the-art entangled photon sources. Additionally, the implementation of the rather complex structure of lenses and beam-splitters is addressed in these studies and radiation effects on single photon detectors have already been investigated in detail \cite{kaiser20081}. These first attempts do show, that a quantum mission based on state-of-the-art technology is feasibly and requires the integration into commercially available space-laser terminals as a next step. 

As outlined above, quantum communication provides a novel way of information transfer. Even though it is still under development, it has the potential to become our future technology for communication and computation. First proposed experiments in space will serve as a very good platform to test these concepts and could pave the way for follow up industrial systems. On a very long term perspective it is highly interesting to test quantum mechanics at distances on the order of millions of km, and even beyond. Furthermore, an ultimate experiment regarding the role of randomness and humans free-will could be performed by two individuals, separated by at least one light second, who each measure entangled particles and separately choose the setting of their analyzer. To extend the scale of quantum mechanical states over astronomical distances might provide us with a suitable insight on the link between gravitation, quantum mechanics and even more. Clearly, these experiments require advances in technology not even foreseeable today. Nevertheless, the proposed experiments are a major step in investigating these fundamental questions as well as enhancing the technology for the society's benefit.

\section{Higher Dimensions}
So far, we have focused only on qubits, which are quantum mechanical two-level systems. This is a natural choice, as all of our classical data storage, transmission, and processing is based on classical two-level systems that encode zeros and ones. There are only a very few exotic exceptions, such as the \textit{Setun} computer build in Soviet union in the late 1950s, which used trinary logic.

However, if one were to look at nature's way of encoding and processing information, one would be surprised to find that it uses a higher-level system: DNA (deoxyribonucleic acid) uses four types of nucleobase (Adenine, Guanine, Cytosine and Thymine) to encode information. Three nucleobase together encode one amino acid, the basis of biological life. If nature---optimized over hundreds of million of years through evolution---uses a higher-level system for encoding information, we see no reason why one shouldn't investigate its use in quantum information as well!

There are two types of high-dimensional systems that depend on whether one considers discrete or continuous parameters. An example of a continuous degree-of-freedom (DoF) is the position (or likewise, the momentum) of a photon. Quantum correlations in this DoF have been used for interesting new types of imaging schemes such as quantum ghost imaging, where the image of the object can only be seen in the correlations of the photons \cite{Pittman:1995jb,Strekalov:1995ub,Malik:2010cr}. A different, even more counterintuitive quantum imaging procedure was recently demonstrated where an object was imaged without ever detecting the photons which were in contact with the imaged object \cite{Lemos:2014gw}. 

In some scenarios, a discrete basis is more advantageous. In classical communications or data storage, for example, information is encoded either as a 0 or a 1; fractional numbers in between are not used. The same is true for quantum communication or quantum computation, even with larger alphabets. A natural basis that uses a discrete DoF of a photon is its orbital angular momentum, which is presented in the next section. Other possible bases can be constructed by the discretisation of continuous parameters such as position or wavelength. 

\subsection{Twisted Photons}
\begin{figure}[h!]
\centering
\includegraphics[scale=0.57]{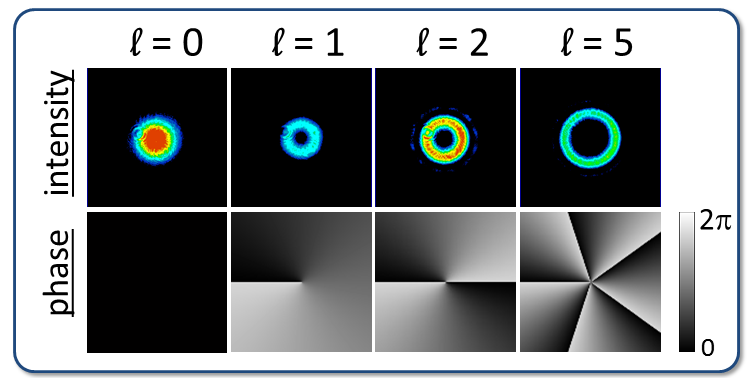}
\caption{Intensities and phase information of orbital angular momentum beams. The intensity is collected with a camera. The OAM=0 mode is the well-known Gaussian distribution. OAM larger than 0 show a ring, or doughnut structure. The lower line shows that these structures have a twisted phase-front, with 2$\pi\ell$ phase-change in a ring. In the center, they have a phase singularity - also known as Vortex. The vortex is the reason why there is no intensity in the center. (Image by Mario Krenn, copyright University of Vienna)}
\label{fig:OAMint}
\end{figure}

If one investigates the spatial profile of a laser beam with a camera, one usually finds that it has a Gaussian shape. However, that is only a special case of a much more complex family of fundamental spatial structures, or modes. One very convenient set of modes are the so-called Laguerre-Gaussian modes \cite{MolinaTerriza:2007ig,allen1999orbital,Yao:2011ve}. In Figure \ref{fig:OAMint}, the intensity and phase structure of a Gaussian mode ($\ell=0$) compared to Laguerre-Gaussian modes ($|\ell|>0$) are shown.

In contrast to its polarisation, which is a property related to its spin angular momentum, a photon with a Laguerre-Gaussian mode structure can also carry orbital angular momentum (OAM). The spin and orbital-angular momenta have distinct physical properties: if a laser beam with circular polarisation illuminates a small particle, the particle will start to rotate around its own axis. However, if a beam with orbital angular momentum shines on a particle, it starts to rotate around the external orbit defined by the laser beam \cite{he1995direct}. Surprisingly, the OAM of photons and its connection to Laguerre-Gauss modes was identified only recently in 1992 \cite{Allen:1992by}.

Interestingly, the OAM quantum number of a photon can theoretically take on any integer number between $-\infty$ and $\infty$. This allows one to encode a huge amount of data onto a single photon \cite{vcelechovsky2007optical,Gibson:2004fw}. In classical communications, this can improve the data rates enormously. Recent experiments have demonstrated data transmission of 100 Tbit/sec by using the OAM of light together with other DoFs \cite{Wang:2012ha,huang2013100}. In quantum communication, secret sharing protocols have been developed that use OAM modes as an alphabet for encoding \cite{Groeblacher:2005ec,MolinaTerriza:2005tr,Walborn:2006jv,Mafu:2013jk,Mirhosseini:2015fy}. Not only do such protocols offer an increased data rate, they also provide an improved level of security against eavesdropping attacks \cite{Wang:2005gk,huber2013weak}.

\subsection{High-Dimensional Entanglement}
Earlier in this chapter, entanglement was explained in the context of photon polarisation, which is a two-level system. In such systems, the separated photon pair can share one bit of information in a non-local manner, referred to as an entangled bit or ``ebit".

\begin{figure}[h!]
\centering
\includegraphics[scale=0.35]{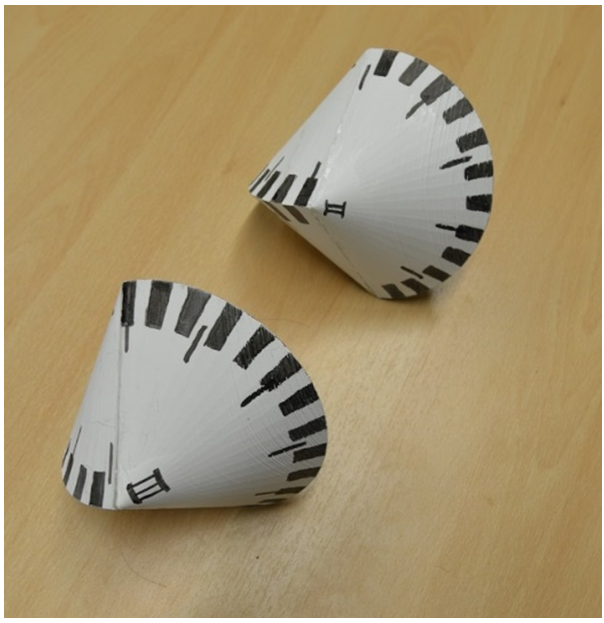}
\caption{Two classical 100-sided dice. If one were to roll them, it is very unlikely that they would both show the same number. However, were they high-dimensionally entangled, they would both always show the same number. Note: such a metaphor for quantum entanglement is limited in that one cannot visualize the results of correlated measurement outcomes in superposition bases. This is key for distinguishing entanglement from classical correlations. (Image by Mario Krenn, copyright University of Vienna)}
\label{fig:100die}
\end{figure}

However, if we consider larger dimensional systems such as the OAM of photons, one can easily imagine that a pair of photons entangled in their OAM could share much more information than photons entangled in their polarisation. Such modes get bigger in size as the OAM quantum number $\ell$ is increased. Thus, the amount of information carried by them is only limited by the size of the optical devices used, or more generally, by the size of the universe itself! A natural question that arises is whether there exists a limit to the amount of information that can be non-locally shared between two entangled photon pairs. This question is being investigated in several laboratories around the world \cite{vaziri2002experimental,torres2003preparation,molina2004triggered,pors2008shannon,agnew2011tomography,Dada:2011vc,romero2012increasing,mclaren2012entangled,salakhutdinov2012full,giovannini2013characterization,Krenn:2014jy,bolduc2015direct}. These efforts have confirmed that two distant photons can be entangled in hundred and more dimensions of their spatial mode structure. This means that by measuring the first photon of the entangled pair, one will observe one definite result out of the hundred possible outcomes. This immediately tells us the outcome of a similar measurement on the second, distant photon. However, the strangeness lies in the fact that the two photons did not have a definite value before they were measured. Only when the first photon is observed does the common state become a reality, and the second photon gets a defined value. 

Photons entangled in their orbital angular momentum also enable the possibility to explore more complex types of entanglement that is not possible with two-dimensional entangled states. Recent state-of-the-art experiments have shown the entanglement of eight photons \cite{Yao:2012fp}, nine superconducting circuits \cite{Kelly:2015gi}, and fourteen ions \cite{Lanyon:2014eh}. However, these experiments have singularly focused on increasing the number of particles entangled, while remaining in a two-dimensional space for each particle. The OAM of light was recently used to create the first entangled state where both, the number of particles and the number of dimensions, was greater than two \cite{malik2015multi}. This state involved three photons asymmetrically entangled in their OAM: two photons resided in a three-dimensional space, while one photon lived in two dimensions. Interestingly, this asymmetric structure only appears when one considers multi-particle entanglement in dimensions greater than two \cite{Huber:2013ie}. Such states also enable a novel ``layered" quantum communication protocol. For example, if three parties were to share the state described above, all three would have access to one bit of secure information, allowing them to generate a secure random key for sharing information. However, part of the time, two of the parties would have access to another bit of secure information. This would allow them to share an additional layer of information unknown to the third party in the communication scheme. This protocol can be generalised to include multiple layers of information shared asymmetrically amongst many different parties.

\subsection{Mutually Unbiased Bases in high dimensions}
Earlier in this chapter we have learned that for 2-dimensional systems, three unbiased bases exist. For larger dimensions, one finds more of these unbiased bases: in 3 dimensions there are 4 bases, in 4 dimensions there are 5 bases. In fact, it is known that for every prime-power dimension (with d=$p^n$), the number of MUBs is (d+1). That means, in dimension d, there are (d+1) different ways to encode information. Now there is one very surprising fact: If the dimension of the space is not a prime-power, it is not known how many MUBs there are. The first of those cases is dimension 2$\cdot$3=6 \cite{bengtsson2006three,wiesniak2011entanglement}. Numerical search has only found 3 MUBs, and it a conjecture that there are only 3 MUBs. It is fascinating because it means that in 5 dimensions, there are more ways to encode information in different ways than in 6 dimensions, even though intuitively one might think that a larger space allows for more ways to embed information in different ways. This is crucial for quantum communication, because the number of MUBs is directly connected to the robustness (against noise and eavesdropping-attacks) of the protocol. The more different ways of encoding the information, the more secure the system is.

\subsection{High-dimensional Quantum Key distribution}

Quantum cryptography based on photons carrying OAM is similar to the schemes developed for polarisation that are explained earlier in this chapter. High-dimensional analogs to the BB84 and Ekert QKD protocols have been developed that use OAM for encoding \cite{Malik:2014ht}. Similar to polarisation-based QKD, OAM-based QKD requires measurements to be performed in mutually unbiased bases to guarantee security against eavesdropping. The earliest such protocol was demonstrated with photons entangled in three dimensions of their OAM ($\ell=0,+1$, and $-1$) \cite{Groeblacher:2005ec}. The high-dimensionally entangled photon pairs were produced in a BBO crystal and sent to two separate stations, where basis transformations were randomly performed by two holograms mounted on moving motorised stages at each station. The photons were then probabilistically split into three paths where their OAM content was measured by three additional holograms. In this manner, a three-dimensional key was generated with an error rate of 10\%. Security was verified by testing for the presence of entanglement via a high-dimensional Bell inequality.

One of the challenges in using OAM modes for quantum communication is the ability to sort single photons carrying OAM. The QKD scheme described above used beam splitters and holograms to projectively measure the OAM content of the single photons. This resulted in a scheme that was photon-inefficient, i.e. only one out of every nine photons was actually used for communication. While techniques for efficiently sorting the OAM of single photons existed, they relied on $N$ cascaded Mach-Zehnder interferometers for sorting $N+1$ OAM modes \cite{Leach:2002wy}. Thus, the use of such a device in a quantum communication scheme was impractical due to issues of complexity and stability. However, in 2010, the group of Miles Padgett developed a refractive device that could sort the OAM of a single photon \cite{Berkhout:2010cb}. This device ``unwrapped" the helical wavefront of an OAM mode, transforming it into a plane wave with a tilted wavefront. The amount of tilt was proportional to the OAM quantum number $\ell$, allowing these modes to be separated by a simple lens. This device provided a diffraction-limited sorting efficiency of 75\%, which was improved to 93\% by the addition of two additional holographic transformations \cite{Mirhosseini:2013em}.

\begin{figure}[h!]
\centering
\includegraphics[scale=0.5]{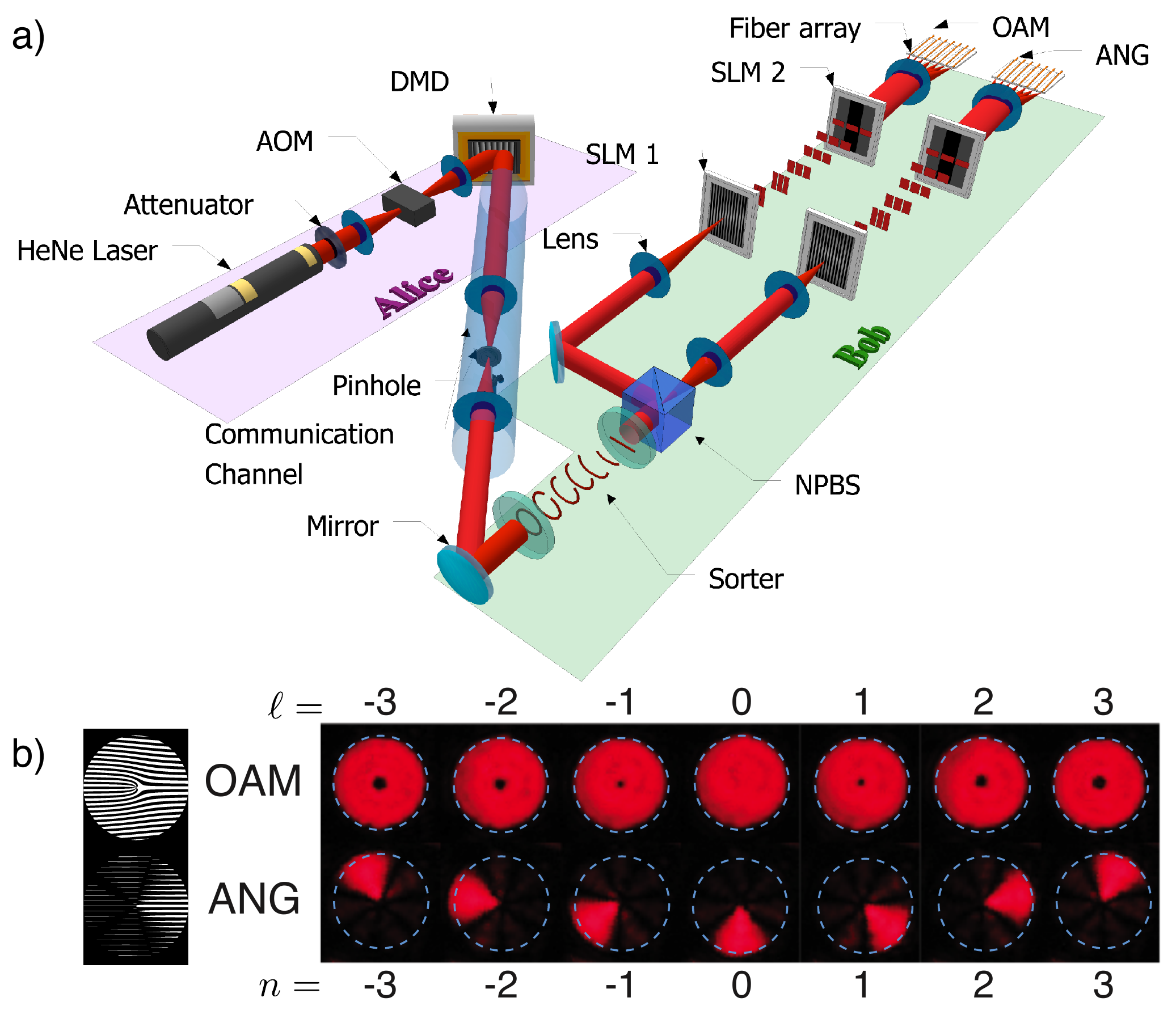}
\caption{(a) An OAM-based BB84 scheme for quantum key distribution. Alice encodes a random key in a seven-dimensional alphabet consisting of OAM modes using a high-speed digital micro-mirror device (DMD). Bob sorts these modes using an OAM sorter and four additional holograms implemented on spatial light modulators (SLMs). Using this scheme, Alice and Bob are able to communication with a channel capacity of 2.05 bits per sifted photon. (b) CCD images showing the intensity profiles of the seven-dimensional alphabet in the OAM basis, as well as the mutually unbiased basis of angular (ANG) modes. Examples of binary holograms for generating these modes are shown on the left (Figure adapted from Ref.~\cite{Mirhosseini:2015fy}).}
\label{fig:OAMQKD}
\end{figure}

The development of this device allowed photon-efficient OAM-based quantum communication schemes to be realized in the laboratory. Recently, a BB84 protocol using a seven-dimensional OAM alphabet was performed which made heavy use of the OAM sorter discussed above \cite{Mirhosseini:2015fy}. Additionally, a digital micro-mirror device (DMD) was used to generate OAM modes at a rate of 4 kHz, which is much faster than the rates attainable with spatial light modulators. The key was encoded in the OAM basis as well as the mutually unbiased of the so-called angular modes (ANG), as shown in Fig.~\ref{fig:OAMQKD}(b). Using this scheme, Alice and Bob were able to communicate securely at a rate of 2.05 bits per sifted photon. Their generated key had an error rate of approximately 10\%, which was below the bounds for security against coherent attacks in a seven-dimensional QKD link. This experiment served as a proof-of-principle demonstration of OAM-based QKD. Several technological improvements (discussed in Ref.~\cite{Mirhosseini:2015fy}) will be required to take such a scheme into the real world.

\subsection{Large Quantum Number Entanglement}
Twisted photons not only allow access to a very large state space, but also give access to very high quantum numbers. Photons can carry $\ell\hbar$ of angular momentum, and $\ell$ can be arbitrarily large. Usually, quantum phenomena are only observed in the microscopic world. Here however, with twisted photons it is possible to create entanglement between photons that differ by a very large amount of angular momentum. Theoretically, there is no upper limit of the number of angular momentum, which would give rise to the possibility of entanglement of macroscopic values of angular momentum.

\begin{figure}[h!]
\centering
\includegraphics[scale=0.7]{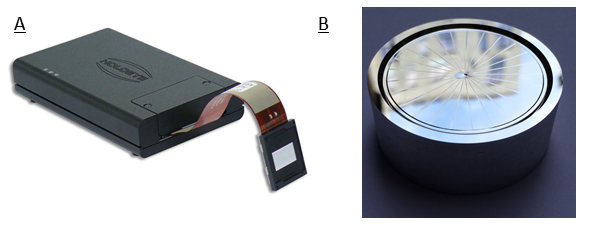}
\caption{Different ways to create photons with large angular momentum. \underline{A}: A spatial light modulator consists of a liquid crystal display. The display consists of roughly 1000x1000 pixels, which perform phase shifts from zero to 2$\pi$. The flexibility allows to create arbitrary phase structure, thus arbitrary structures of the modulated light. However, due to their finite resolution, there is an upper limit of roughly 300$\hbar$. (Image by HOLOEYE Photonics AG) \underline{B}: A different method that can create angular momentum of up to 10.000$\hbar$ are fixed phase holograms build out of aluminium. In compensation for the lower flexibility, the holograms can be produced very precise, which is responsible for the much larger possible angular momentum. (Image by Robert Fickler, copyright University of Vienna)}
\label{fig:OAMholo}
\end{figure}

With this method, it was possible to show that two photons with a difference of 600$\hbar$ can be entangled \cite{Fickler:2012hj}. If the first photon carries 300$\hbar$ of angular momentum, the second carried -300$\hbar$, and vice versa. While being entangled in an two-dimensional subspace, it was the largest quantum number difference achieved. In that experiment, a spatial light modulator has been used, which can be seen in Fig.~\ref{fig:OAMholo}. Recently, using novel methods to encode very large angular momentum at single photons, it was able to show entanglement of photons with a quantum number difference of 10,000$\hbar$.

An important question that needs to be answered is the definition of \textit{macroscopic angular momentum}, and which phenomena might arise from that. For example, there are predictions that photons close to a black hole change their angular momentum \cite{tamburini2011twisting}. As black holes are purely general relativistic objects, and entanglement is a purely quantum mechanical phenomenon, a deeper investigation into of these effects will be exciting.

\subsection{Long-distance transmission of twisted photons}
In a quantum communication scenario, the encoded information needs to be distributed between two parties. Usually one would think that optical fibers are the ideal solutions. Unfortunately, the information in twisted photons is not conserved in propagation through conventional fibers: Different modes mix in fibers, therefor the output is different than the input. Although recent advances show that special fibers can be used to transmit the first higher-order OAM modes for more than one kilometer \cite{bozinovic2013terabit}, and reach a classical communication rate in the order of Terabit, this technology is still in its infancy. Specifically, it hasn't been used in the realm of quantum physics yet. An alternative method is the transmission through free-space. In the case of earth-to-satellite quantum communication, this is the only possibility in any case.

\begin{figure}[h!]
\centering
\includegraphics[scale=0.35]{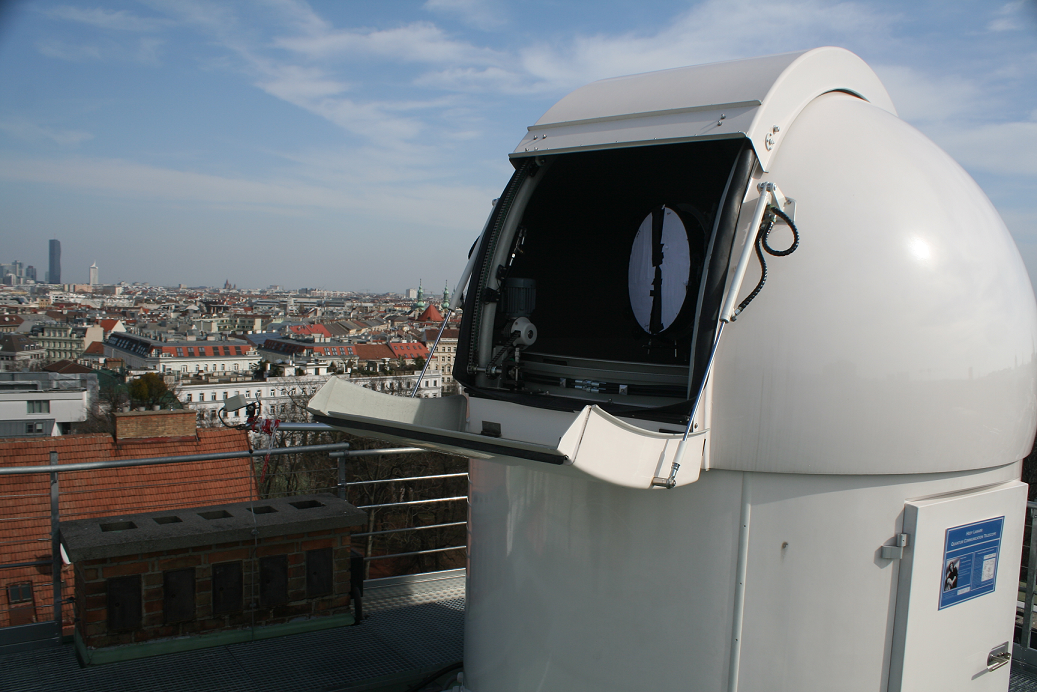}
\caption{Receiver at the Hedy Lamarr Quantum Communication Telescope for the first free-space long-distance entanglement distribution experiment with a high-dimensional degree of freedom. (Image by Robert Fickler, copyright University of Vienna)}
\label{fig:kuppel}
\end{figure}

If long-distance transmission is considered, immediately the influence of atmospheric turbulence has to be taken into account. Varying pressure and temperature influence the structure of twisted photons. The question is: How much? While many mathematical and lab-scale studies have been performed, experimental investigation of that question are rare. Only recently, the first classical \cite{Krenn:2014bx,lavery2015study,krenn2016twisted} and quantum communication \cite{krenn2015twisted} experiments have been performed over free-space intra-city link of more than 1 kilometer distance. Those results show that quantum entanglement with twisted photons can be distributed over larger distances, and the quality can be improved with technology that is already implemented in lab-scale experiments \cite{rodenburg2014simulating,ren2014adaptive,xie2015phase}. As such, it could be a reliable way to distribute high-dimensional entanglement in a future quantum network.

%\subsection{Continuous Variables}%
%entanglement in position-momentum, discretisation---pixel entanglement, 4D QKD based on position-momentum, high-dim entanglement with pixel arrays and EMCCDs, limitations, briefly mention ghost imaging and mandel (not really "communication")%

\section{Conclusion}
The possibility to share secret messages is of utmost importance for our society. From simple things like sending emails which can't be read by an eavesdropper to the transmission of highly sensitive information between governments that needs to be secure for decades---cryptography plays a key role in ensuring privacy, economic stability, and stable relations between countries worldwide.

As we have seen, classical cryptographic systems are vulnerable to various types of eavesdropping attacks. The problem is that either the secret key needs to be transmitted over insecure channels, or (in a public-private cryptography system) the security relies on mathematical conjectures that specific properties are difficult to calculate. Furthermore, quantum computing algorithms can significantly reduce the required time to find solutions for such problems (finding prime factors of large numbers, or calculating a discrete logarithm). On top of all this, back-doors can be implemented into these algorithms such that they perform as expected, but the creator of the algorithm obtains additional information. Such attacks have been widely discussed in connection with a weak generator for pseudo random numbers certified by NIST \cite{shumow2007possibility, perlroth2013government}.

The need for overcoming these problems posed by classical asymmetric cryptographic systems has led to the development of a field called Post-Quantum-Cryptography. There, problems which are believed to be more difficult than factoring large numbers are used to prepare a public and private key. Such methods are not practically used yet because of performance issues and unclear results on their security. While there are no classical or quantum algorithms to solve such problems yet, it is only conjectured that they are difficult to solve---a breakthrough in (quantum) complexity theory or novel kind of computations might only shift the problem into the future.

The only unconditionally secure encryption requires a random key with the same size as the message, a so-called one-time pad. The question is, how can such a key be distributed securely? Quantum key distribution provides a solution to that question, by exploiting quantum mechanical properties of individual particles. Several newly founded companies already provide small-scale quantum key distribution systems, such as ID Quantique in Switzerland, MagiQ Technologies in USA, QuintessenceLabs in Australia or SeQureNet in France.

As shown in this chapter, fundamental investigations test the feasibility of global quantum networks, on the order of 100 kilometers on the Earth's surface, as well as between ground and space. A second path of research focuses on more complex quantum states, to improve data-rates and robustness against noise and eavesdropping attacks. The experiments discussed in this chapter form only a small subset of experimental efforts currently in progress around the world. It is clear that we are perched on the edge of a quantum communication revolution that will change information security and how we understand privacy for years to come. 

\subsection*{Acknowlegdements}
We acknowledge cooperation with Jian-Wei Pan and the Chinese Academy of Sciences. This work was supported by the European Space Agency, the European Research Council (ERC Advanced Grant No. 227844 ``QIT4QAD" and SIQS Grant No. 600645 EU-FP7-ICT), the European Commission (Marie Curie grant ``OAMGHZ"), the Austrian Science Fund (FWF), the Austrian Academy of Sciences (\"OAW) and the Austrian Research Promotion Agency (FFG) within the ASAP program from the Federal Ministry of Science and Research (BMWF), as well as the John Templeton Foundation. 

\bibliographystyle{spbasic}
\bibliography{references,all}
\end{document}